# A 3GPP Perspective on Spectrum Sharing for 5G-to-6G Migration: From LTE-NR DSS Lessons to MRSS Design

Xingqin Lin

*Abstract*—**Dynamic spectrum sharing (DSS) played an important role in the 4G-to-5G transition by allowing 5G new radio (NR) to reuse legacy spectrum without immediate static refarming. Yet practical deployments also exposed the cost of coexistence of NR with long-term evolution (LTE), including overheads, control-channel bottlenecks, neighbor-cell interference, etc. As 6G begins to take shape, spectrum scarcity below 7 GHz is again making 5G-6G spectrum sharing a migration tool of interest. Multi radio access technology spectrum sharing (MRSS) is being considered by the 3rd generation partnership project (3GPP) as a key mechanism for 5G-6G coexistence. This article reviews the lessons learned from LTE-NR DSS and examines how those lessons should shape MRSS design. The main challenge is no longer basic coexistence feasibility, but coexistence efficiency which determines whether MRSS will become a broadly usable framework for 5G-to-6G spectrum migration.**

## I. INTRODUCTION

The transition to a new mobile generation has always been shaped by spectrum, but the spectrum problem confronting 6G is more constrained than in earlier generations [1]. In previous transitions, operators could often rely on either newly identified spectrum or a temporary separation between legacy and emerging systems across different bands. For 6G, that separation will be much harder to maintain. Wide-area coverage, mobility robustness, and reliable indoor service will continue to depend heavily on frequency range 1 (FR1) spectrum below 7 GHz, yet the most valuable portions of that spectrum are already occupied by existing cellular systems. As a result, the key 6G migration question is how to introduce 6G into spectrum that remains heavily used by 5G new radio (NR) [2].

This coexistence pressure is especially relevant for FR1 spectrum below 7 GHz because these bands remain important for wide-area coverage, mobility robustness, and indoor reach. Although spectrum in the 7–15 GHz range and millimeter-wave bands can provide substantial additional capacity, such bands are unlikely to fully replace low-band and mid-band FR1 as coverage anchors during the early 6G phase [3]. Therefore, current 6G migration discussions in 3GPP consider the reuse of spectrum already occupied by 5G NR as an important deployment scenario, rather than assuming that 6G can rely only on newly identified spectrum [2]. In this sense, the industry is returning to a familiar question: how can a new radio access technology (RAT) be introduced into a band that still remains operationally and commercially important for the incumbent RAT?

The most relevant historical precedent is dynamic spectrum sharing (DSS) between long-term evolution (LTE) and NR [4]. DSS was introduced to solve a practical deployment problem: operators wanted to launch 5G in coverage-relevant bands without shutting down LTE too early and without waiting for significant new low-band or mid-band spectrum. By allowing LTE and NR to coexist in the same carrier, DSS provided a soft-refarming mechanism that helped accelerate 5G introduction. However, the same experience also exposed a central lesson that is highly relevant for 5G-to-6G migration: the success of a sharing mechanism cannot be judged only by whether coexistence is technically possible. It must also be judged by how much useful data resource remains after protection of incumbent structures, how much control flexibility remains, how robust the framework is to interference in multi-cell environments, and whether it continues to make sense as the installed base gradually shifts from the incumbent RAT to the new RAT.

These lessons are precisely why multi-RAT spectrum sharing (MRSS) has become a focal point in current 6G migration discussion in 3GPP [2][5]. Compared to DSS, MRSS is being positioned as a more mature coexistence framework built on a better technical foundation [6][7]. The reason is that 5G NR is a much leaner incumbent RAT than LTE was. It does not impose LTE-style always-on cell specific reference signal (CRS), its reference signals and control resources are more configurable, and many of its physical-layer design elements are already closer to what a future 6G RAT would naturally want to align with. This means that 6G can, in principle, be designed from the outset to coexist with a leaner and more configurable incumbent 5G RAT, which, unlike LTE, does not rely on always-on CRS and has more configurable control and reference-signal structures. That is the main source of optimism around MRSS. At the same time, a better coexistence baseline does not guarantee an efficient migration framework. For MRSS, the central question is not only whether 5G and 6G can share spectrum efficiently, but whether 6G can deliver sufficiently compelling new value to justify early deployment. In other words, MRSS becomes attractive only if it enables operators to introduce promising 6G capabilities, such as improved performance, new service support, or more efficient network operation, without waiting for complete 5G spectrum refarming[8].

Developing an efficient MRSS framework requires two linked steps. The first is to revisit DSS carefully as a case study to draw lessons. The second is to examine why MRSS starts from a stronger position, while also identifying the issues that could still prevent that stronger position from translating into an efficient spectrum sharing framework. Therefore, the rest of this article is organized as follows. We first discuss 5G-to-6G migration path. Then we review the major lessons from DSS and explain why MRSS is technically better positioned than DSS. After that, we discuss the main technical foundations of MRSS, followed by concluding remarks.

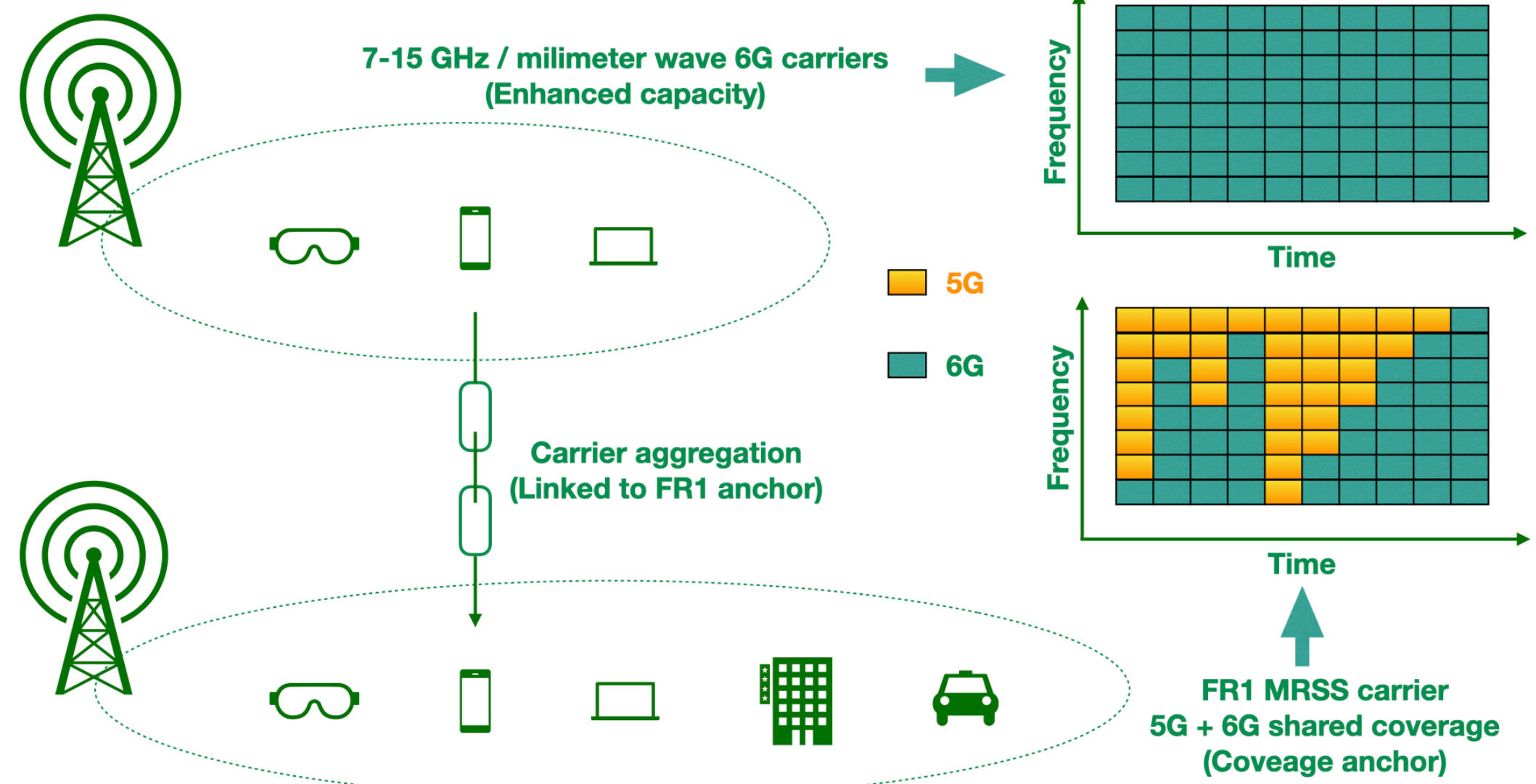

**Figure 1: An illustration of 5G-to-6G migration with MRSS and 6G carrier aggregation.**

## II. 5G-TO-6G MIGRATION PATH

The most credible deployment path for 6G is not one in which a completely new network suddenly replaces 5G across all relevant bands, but one in which 6G is introduced first where it can create immediate value while continuing to rely on existing spectrum assets and site infrastructure. This deployment logic is not completely new; a similar motivation existed during the 4G-to-5G transition. The important difference is the technical and architectural starting point. In LTE-NR DSS, NR had to coexist with an incumbent LTE carrier containing dense CRS and relatively rigid control-channel structures. In 5G-to-6G MRSS, the incumbent RAT is 5G NR, which is more configurable and has no LTE-style always-on CRS. In addition, 6G migration is being considered with the hindsight of DSS limitations, with stronger emphasis on first-release coexistence design and standalone operation. In practical terms, this means that wide-area early deployment will continue to depend heavily on FR1, even if 6G later expands aggressively into 7-15 GHz and millimeter wave frequencies. As a result, the migration problem becomes one of how to introduce 6G in those bands without prematurely shutting down 5G or accepting the efficiency loss of blunt static partitioning. This is where MRSS becomes central. In particular, 3GPP has agreed that the 6G radio access network shall support MRSS between 6G radio and 5G NR [2].

This migration path has several advantages. First, it minimizes the need for new site acquisition and extensive new hardware deployment during the earliest phase of 6G introduction. Reusing existing FR1 spectrum and radio infrastructure is attractive because it allows 6G rollout to be driven more by software and standardization readiness than by waiting for new nationwide spectrum assignments or large-scale rebuilds. Second, MRSS gives operators a path to extend 6G coverage beyond isolated capacity hotspots once the value of early 6G services has been validated. Operators may still begin with targeted deployments in selected high-traffic areas, but an MRSS-capable FR1 carrier can reduce the barrier to broader coverage expansion because it avoids waiting for complete refarming of 5G spectrum. Third, it naturally matches the service-layer reality that wide-area connectivity and high-capacity bursts do not need to originate from the same band. In such a system, an MRSS FR1 carrier acts as the coverage anchor, while additional carriers in higher bands provide enhanced capacity performance, as illustrated in Figure 1.

From a technical standpoint, this migration model is tightly coupled with carrier aggregation. If the shared FR1 carrier is to avoid becoming the only meaningful 6G resource, then it must be possible to aggregate it efficiently with additional carriers. This is why current migration discussion in 3GPP frequently treat MRSS and 6G carrier aggregation as related topics. The implication is that the shared carrier should be designed as part of a broader multi-carrier system. That affects both downlink and uplink design. On the downlink side, control signaling and scheduling coordination must support traffic and mobility across multiple carriers. On the uplink side, the FR1 anchor may need to carry uplink control information (UCI) associated with additional downlink carriers.

A related aspect of the migration path is architecture simplification. One recurring migration lesson is that multiple intermediate architecture options can create uncertainty, complicate ecosystem alignment, and delay convergence on the most practical deployment model. For this reason, the migration favors a single-step path with standalone 6G, avoiding non-standalone deployment in 5G [2]. In such a case, MRSS can remove one important spectrum barrier for standalone 6G by allowing 6G radio access to operate in shared FR1 spectrum before full refarming. However, MRSS by itself is not sufficient for standalone 6G deployment. Thus, MRSS should be understood as a spectrum-enabling mechanism within a broader standalone 6G deployment framework. This implies that MRSS should be designed to work under a standalone 6G architecture while avoiding requirements that force changes to incumbent 5G user equipment (UE).

A further migration consideration is multi-generation coexistence, especially with long-lived internet-of-things (IoT) systems. Although the primary motivation for MRSS is broadband 5G-6G spectrum reuse and the use of FR1 as an anchor for aggregation with higher-frequency capacity carriers, practical FR1 deployments may also include long-lived IoT systems. This is especially relevant in sub-GHz coverage bands, where narrowband IoT (NB-IoT) and LTE machine type communication (LTE-M) devices can remain in service for long lifecycles [9][10]. Therefore, NB-IoT/LTE-M coexistence should be treated as a deployment boundary condition rather than the main motivation for MRSS. In such cases, semi-static protection of a small set of IoT resources can coexist with dynamic 5G-6G sharing over the remaining carrier. Indeed, 3GPP has agreed that the 6G RAT shall support coexistence with NB-IoT and LTE-M via semi-static configuration [2].

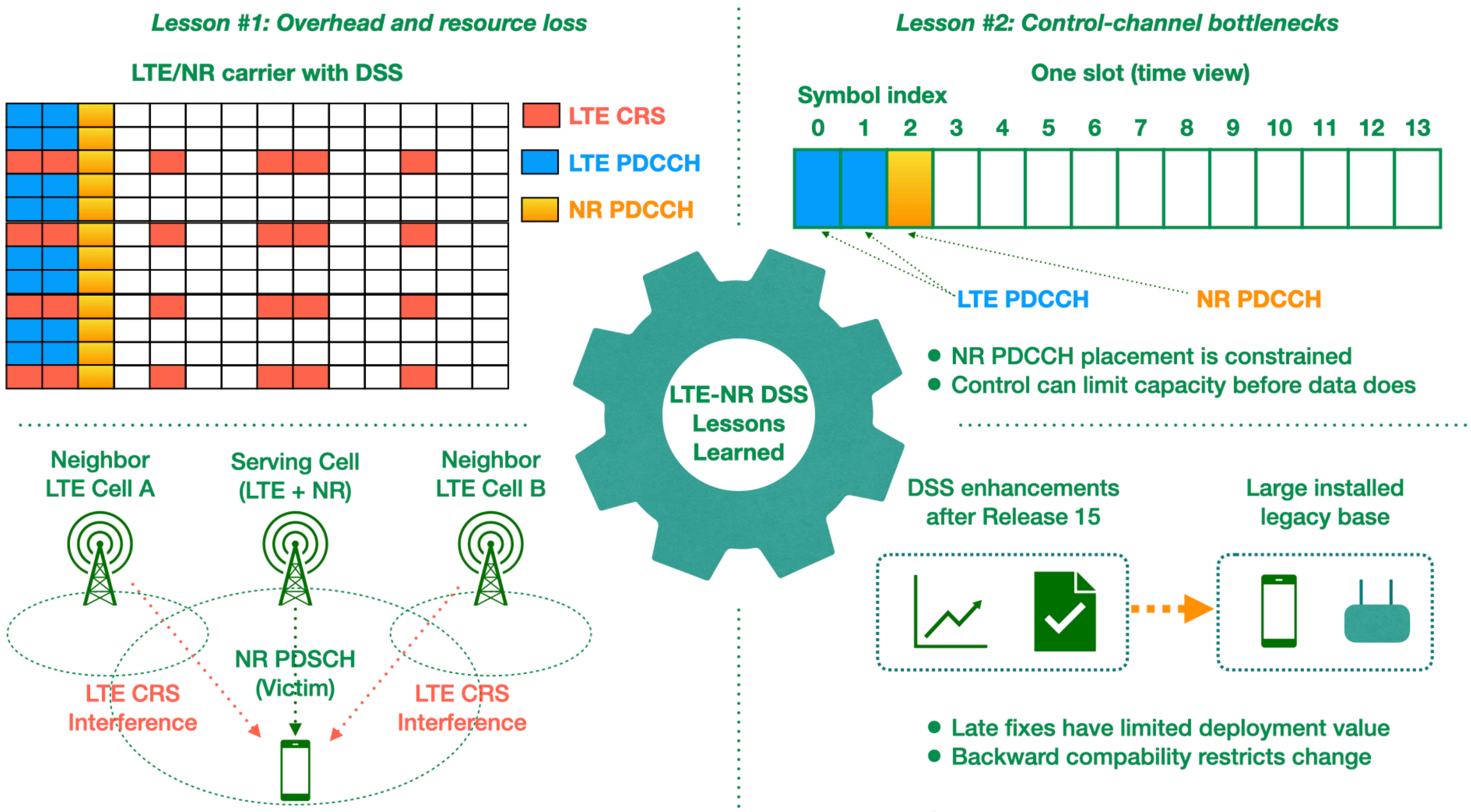

**Figure 2: An illustration of the main lessons learned from LTE-NR DSS.**

## III. LESSONS FROM DSS

At a high level, the main practical DSS mechanisms fell into three categories [6]. The first used LTE multicast-broadcast single-frequency network (MBSFN) subframes, where LTE data is muted after the control region, leaving a block of OFDM symbols available for NR use. This approach was especially convenient for carrying NR signals that could not tolerate LTE CRS collisions, such as synchronization signal block (SSB) transmissions. The second relied on mini-slot based scheduling, allowing NR to use only subsets of symbols that avoided LTE structures. The third, and ultimately the most important for regular data transmission, used rate matching around LTE CRS so that the NR scheduler and UE would treat certain resource elements (REs) as unavailable [12]. These options reflected different tradeoffs between simplicity and spectral efficiency. MBSFN made coexistence more transparent to legacy LTE devices but could reduce LTE user throughput substantially if used too frequently. Mini-slot scheduling avoided some collisions but came with its own overheads, especially from shortened transport blocks and additional demodulation reference signal (DMRS) burden. Rate matching was the most spectrally attractive method for NR data, but it also exposed the limits of how well a new RAT can coexist around dense incumbent structures. In this section, we summarize four main lessons learnt from LTE-NR DSS, as illustrated in Figure 2.

**Lesson #1: Overhead and resource loss**. The first major lesson from DSS is that coexistence overhead accumulates quickly once all incumbent structures are accounted for. LTE physical downlink control channel (PDCCH) typically occupies the first one to three symbols of the subframe, leaving fewer symbols available for NR data and control. LTE CRS then occupies sparse but recurring RE patterns across the bandwidth, requiring NR physical downlink shared channel (PDCCH) to avoid them. NR SSB signals often require additional protection, which in many practical implementations meant using MBSFN subframes. LTE synchronization signals and physical broadcast channel (PBCH) can also force additional avoidance in specific time-frequency regions. The result is that the resources that are actually usable for NR payload transmission in DSS are much fewer than in a pure NR carrier. For example, consider a configuration with two NR DMRS symbols, two LTE PDCCH symbols, and one NR PDCCH symbol in each slot. Table 1 compares the useful RE budget per physical resource block (PRB) on an LTE-NR DSS carrier across different LTE CRS port configurations. The values in Table 1 are obtained by RE accounting over one PRB and one 14-symbol slot, i.e., $12 \times 14 = 168$ REs per PRB. For example, with four LTE CRS ports, the DSS carrier reserves 2 LTE PDCCH symbols, 1 NR PDCCH symbol, 2 NR DMRS symbols, and 16 LTE CRS REs outside PDCCH region per PRB. The remaining NR PDSCH REs are therefore $168 - 2 \times 12 - 1 \times 12 - 2 \times 12 -$

| No. of LTE CRS ports | No. of NR PDSCH REs on DSS carrier | No. of NR PDSCH REs on NR carrier | No. of NR PDSCH REs on LTE carrier | NR DSS loss vs. NR carrier | NR DSS loss vs. LTE carrier |
|---|---|---|---|---|---|
| **1** | 102 | 132 | 138 | 22.73% | 26.09% |
| **2** | 96 | 132 | 132 | 27.27% | 27.27% |
| **4** | 92 | 132 | 128 | 30.30% | 28.13% |

Assumptions:

- DSS carrier: 2 NR DMRS symbols, 2 LTE PDCCH symbols, and 1 NR PDCCH symbol per slot
- NR carrier: 2 NR DMRS symbols and 1 NR PDCCH symbol per slot
- LTE carrier: 2 LTE PDCCH symbols per slot

**Table 1: RE loss per PRB in LTE-NR DSS under different LTE CRS port configurations.**

$16 = 92$. By comparison, the pure NR carrier has $168 - 1 \times 12 - 2 \times 12 = 132$ NR PDSCH REs, while the LTE carrier has $168 - 2 \times 12 - 16 = 128$ PDSCH REs. Thus, the DSS loss is $(132 - 92)/132 = 30.30\%$ relative to pure NR and $(128 - 92)/128 = 28.13\%$ relative to LTE. An important observation is that the DSS loss relative to pure NR increases with the number of LTE CRS ports, reflecting the growing coexistence burden as the incumbent LTE reference signal density becomes higher.

**Lesson #2: Control-channel bottlenecks**. The second major lesson from DSS is that control can become more limiting than data. Data channels can often be adapted by puncturing, rate matching, or moving resources in time and frequency. Control channels have far less freedom. Their placement is constrained by PDCCH search space design, monitoring assumptions, aggregation-level requirements, and the need for timely delivery of scheduling grants. When the incumbent RAT occupies the earliest symbols and imposes additional reference signal collisions, the available control region for the new RAT can become very narrow. This happened in DSS. In a common DSS configuration, the beginning of the slot is effectively partitioned into two LTE control symbols plus one NR control symbol. The first two OFDM symbols are reserved for LTE PDCCH so that legacy LTE operation remains unaffected, while the earliest practical location for NR PDCCH is then pushed to the third OFDM symbol. In deployments with four LTE CRS ports, LTE CRS also occupies REs in early symbols, which further reinforces this arrangement and makes it difficult to expand NR control deeper into the slot without creating additional coexistence constraints. The resulting NR PDCCH capacity was limited to a small number of control channel elements (CCEs). That limited PDCCH capacity had to handle both downlink and uplink scheduling and also support common search space signaling, which made the control bottleneck a real operational issue. Control overhead in the uplink could also become problematic, especially when periodic channel state information (CSI) reporting consumed reserved physical uplink control channel (PUCCH) resources on scarce low-band anchors.

**Lesson #3: Neighbor-cell interference**. The third major lesson from DSS is that structured neighbor-cell interference can limit performance even when the serving-cell coexistence mechanism appears well designed. In the simplest DSS view, the serving cell knows the co-located LTE configuration and can rate match NR PDSCH around the relevant LTE CRS pattern. But that assumption breaks down in multi-cell environments, because neighboring LTE cells may introduce their own CRS patterns and physical cell ID dependent shifts. The consequence is that some REs that appear usable from the serving-cell perspective are actually subject to structured LTE interference from neighboring cells, especially in coverage-overlap regions. This interference could degrade NR downlink throughput significantly and motivated several categories of mitigation [11]. One was to extend rate-matching signaling so that the UE could also avoid patterns corresponding to neighboring LTE cells, not just the co-located serving cell. Another was to use broader symbol-level or zero power CSI reference signal (CSI-RS) marking to declare larger classes of REs unavailable whenever the precise neighbor-cell CRS pattern was unknown. A third was to rely more heavily on advanced UE-side receivers, particularly interference cancellation, so that the receiver could estimate and suppress or subtract the interfering LTE CRS. These approaches have different costs. UE-side cancellation requires the UE to detect neighboring LTE cells, estimate their channels, infer interfering RE positions, and rebuild interference. Network-assisted approaches, by contrast, reduce receiver burden but may sacrifice additional REs or symbols if the configured avoidance is broad.

**Lesson #4: Compatibility, timing, and migration flexibility**. A fourth lesson from DSS concerns the timing of improvements and the difficulty of fixing a migration framework once the first device wave has already shaped the ecosystem. Several of the practical limitations of DSS were recognized and addressed, at least partially, after 3GPP Release 15 [13]. Examples include richer rate matching options, interference mitigation capabilities, and mechanisms to ease control restrictions. Yet the adoption of these mitigations remained limited because early Release-15 devices constrained what operators and vendors could count on in practice. The result was that even technically sound improvements could arrive too late to change the field reality. This is perhaps the most important non-physical-layer lesson DSS offers to MRSS. A migration framework that is incomplete in first release may remain constrained for years, even if later releases improve it substantially on paper.

## IV. WHY MRSS IS DIFFERENT

MRSS is often introduced as the natural successor to DSS, but the incumbent RAT is fundamentally different. DSS had to insert NR into a carrier dominated by LTE with rigid time-frequency patterns. In MRSS, by contrast, the incumbent system is 5G NR, which is much leaner and much more configurable than LTE [14]. The key differences between DSS and MRSS are summarized in Table 2 and discussed below.

| Difference | DSS | MRSS | Implication |
|---|---|---|---|
| **Incumbent signal density** | LTE has persistent CRS across the carrier. | 5G has no LTE-style always-on CRS; SSB, PDCCH, CSI-RS, and system info are sparser and more configurable. | Lower coexistence burden in MRSS. |
| **Control configurability** | NR control must fit around rigid LTE control placement. | 5G/6G control can use configurable CORESETs and search spaces, with overlapping, partial, or separate control options. | Control coexistence is more flexible in MRSS. |
| **Migration scope and duplexing** | DSS was mainly associated with FDD coverage bands. | MRSS is relevant to both FDD and TDD, including mid-band TDD. | MRSS supports broader deployment scenarios. |
| **Architecture coupling** | DSS mainly enabled early 5G coverage in legacy spectrum. | MRSS is closely tied to 6G carrier aggregation: FR1 for coverage, higher bands for capacity. | The shared carrier becomes a key coverage and control anchor. |
| **Role of hindsight** | DSS was developed without prior large-scale coexistence experience. | MRSS builds on DSS lessons in control, initial access, signal awareness, and feature timing. | MRSS can be designed more deliberately from the start. |

**Table 2: Key Differences Between DSS and MRSS**

**Difference #1: Incumbent signal density**. LTE imposed persistent CRS across the carrier. In the MRSS case, there is no LTE-style always-on CRS in the incumbent 5G carrier. This removes one of the heaviest coexistence penalties from the DSS experience. The 5G downlink still contains signals and channels that matter for coexistence, including SSB, PDCCH, CSI-RS, and system information, but these structures are much sparser, more configurable, and more schedulable than LTE CRS.

**Difference #2: Control configurability**. LTE PDCCH occupied the earliest symbols of the subframe with limited configurability. By contrast, 5G control is already built around configurable control resource sets (CORESETs) and search spaces [15]. Instead of control being rigidly inherited from a legacy system, MRSS can consider multiple coexistence models for control from the start: fully overlapping control regions, partially overlapping regions, or separate regions with tradeoffs.

**Difference #3: Migration scope and duplexing relevance**. Practical DSS experience was mainly associated with frequency division duplex (FDD) coverage bands, especially because those bands were important for early nationwide 5G coverage and because LTE already occupied them extensively. MRSS may be needed not only in traditional low-band FDD deployments but also in important mid-band time division duplex (TDD) deployments. MRSS, therefore, is being considered from the outset as a framework for both FDD and TDD. The latter introduces different slot structure considerations and different implications for uplink/downlink timing.

**Difference #4: Architecture coupling, especially the link between MRSS and 6G carrier aggregation**. DSS was often discussed as a way to get early 5G coverage in legacy spectrum. For MRSS, one of the expected rollout models is that FR1 shared carriers provide wide-area coverage and early accessibility, while additional higher-band carriers are aggregated to provide enhanced capacity, rather than resorting to dual connectivity. From a system perspective, this architecture raises the importance of control and uplink design on the shared carrier, because that carrier may also need to carry the UCI associated with aggregated downlink capacity.

**Difference #5: Role of hindsight**. DSS was developed without the benefit of large-scale prior experience in dynamic inter-generation sharing between a dense incumbent RAT and a more flexible new RAT. MRSS starts with explicit lessons from DSS about what hurts most in practice. MRSS can be designed with a stronger awareness that coexistence efficiency depends not just on shared data scheduling, but on control multiplexing, initial access, signal awareness, and the timing of feature availability. The value of hindsight also changes the standardization philosophy in 3GPP: MRSS needs to be treated as a first-release migration framework, not as a feature that can be incrementally enhanced in later releases.

In short, MRSS is different from DSS because it offers a cleaner foundation, a broader migration scope, and a more informed design process. But it still requires deliberate technical decisions to translate those advantages into an efficient spectrum sharing framework, which is discussed in the next section.

## V. MRSS DESIGN FOUNDATIONS

The central MRSS design objective is to construct a common time-frequency and control framework that minimizes overhead, limits scheduling restrictions, and keeps the shared carrier close to the efficiency of a pure 6G carrier as device penetration evolves. DSS taught that even technically sound improvements introduced in later releases may have limited field impact if early devices, early networks, and early deployment assumptions have already constrained what can be adopted. Therefore, the first 6G release needs to contain enough of the core MRSS framework that both infrastructure and devices can enter the market with a usable, scalable coexistence model. The resulting design problem spans waveform and numerology, initial access, downlink and uplink control multiplexing, reference signals and interference awareness, as illustrated in Figure 3.

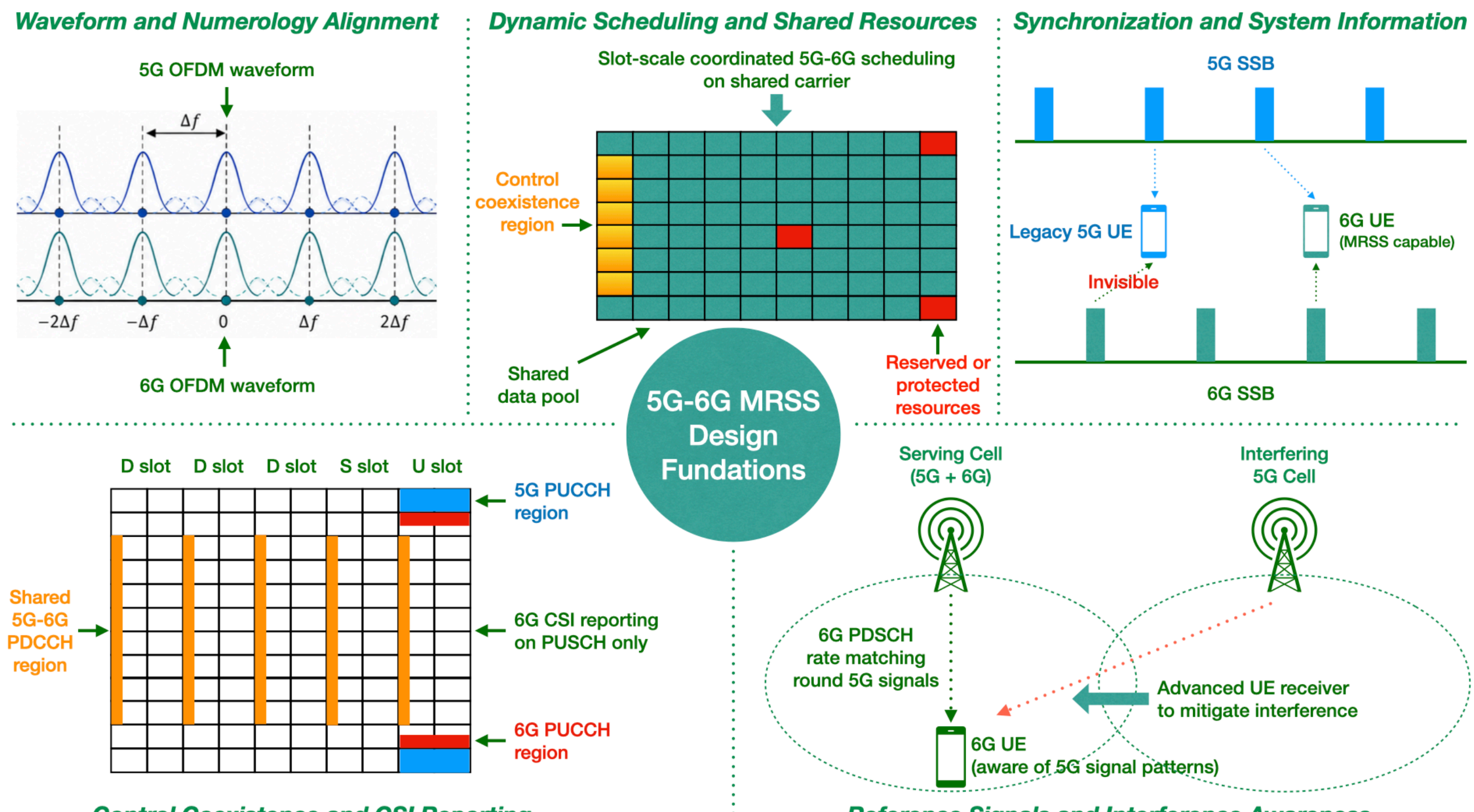

**Figure 3: An illustration of the design foundations of 5G-6G MRSS.**

### A. *Waveform and Numerology Alignment*

The fundamental foundation of MRSS is common waveform structure. The same orthogonal frequency division multiplexing (OFDM) waveform family and aligned numerology are expected to be used by 6G in bands shared with 5G, with aligned subcarriers and compatible channel raster design. The importance of numerology alignment is especially clear in the comparison with DSS rate matching. In DSS, some coexistence options were only practical in 15 kHz subcarrier spacing (SCS) configurations because LTE and NR aligned most naturally there, while broader symbol-level avoidance had to be used when numerology diverged. The lesson carried into MRSS is that aligned subcarriers and compatible slot structures are prerequisites for fine-grained sharing with low overhead. If 5G and 6G are aligned at the PRB and symbol level, then shared scheduling can remain dynamic, rather than forcing coarse muting.

A related issue is the channel raster and carrier positioning framework. Even if 6G evolves some aspects of synchronization and channel discovery relative to 5G, those choices should remain compatible with 5G raster where shared operation is expected. If not, the coexistence efficiency gained from common numerology could be partly lost at the carrier-placement level. In practical terms, this means that MRSS should avoid introducing a channelization design that forces misaligned offsets, extra guard bands, or unnecessary separation between 5G and 6G resources inside the same carrier.

### B. *Dynamic Scheduling and Shared Resources*

Dynamic scheduler based sharing is another foundation of MRSS. In a shared carrier, when one RAT has little or no traffic, the other RAT should be able to use nearly the entire carrier. That means resource sharing cannot be static or semi-static over large regions but must be implemented primarily through coordinated scheduling on a millisecond timescale. To make this work, the schedulers of the two RATs must operate over a common understanding of which resources are dynamically shareable and which resources are reserved or restricted. This resource classification should not be interpreted as a return to LTE-like rigidity or as a direct reuse of DSS. In DSS, the sharing framework was largely shaped by fixed LTE structures, especially CRS and early-symbol LTE PDCCH, which created persistent resource holes and control restrictions. In MRSS, the protected resources are expected to be much sparser and more configurable. The common understanding between 5G and 6G schedulers can therefore be built from a small semi-static baseline for signals that cannot move, combined with dynamic scheduling coordination for data and control resources. This is different from DSS: the objective is not to force 6G around a dense incumbent resource map, but to define coexistence primitives that preserve the configurability of 5G NR while allowing 6G to avoid only the resources that truly require protection.

In the simplest form, the shared carrier can be thought of as having three categories of resources. The first category is the shared data pool, where both RATs can be scheduled dynamically and which should occupy the majority of the carrier in an efficient MRSS design. The second category is reserved or protected resources, which include synchronization, initial access, or other incumbent structures that cannot be dynamically repurposed. The third category is the control coexistence region, which may be fully shared, partially overlapping, or separate depending on the final design. A well-designed MRSS carrier is therefore one in which the shared data pool is maximized, the reserved resources are sparse and predictable, and the control region is handled in a way that avoids excessive overhead. The shared data pool also needs to be governed by QoS- and priority-aware scheduling. If the incumbent 5G carrier supports priority or mission-critical services, the scheduler can protect such traffic through, e.g., admission control and dynamic prioritization. The corresponding 5G resources would be treated as temporarily unavailable to 6G when the priority condition is active, while remaining dynamically shareable when such traffic is absent. In this way, MRSS can support priority 5G traffic without permanently partitioning a large fraction of the carrier.

| Signal / channel | Configuration | Overhead RE in 20 ms | Overhead vs. total RE (%) | Overhead vs. total downlink RE (%) |
|---|---|---|---|---|
| SSB | 4 beams, 20 PRBs, 4 symbols | 3,840 | 0.21% | 0.31% |
| CORESET 0 / common PDCCH | 4 beams, 48 PRBs, 2 symbols | 4,608 | 0.25% | 0.37% |
| SIB1 | 4 beams, 24 PRBs, 4 symbols | 4,608 | 0.25% | 0.37% |
| **CORESET 1 / regular PDCCH** | **270 PRBs, 2 symbols, 32 DL-bearing slots** | **207,360** | **11.30%** | **16.48%** |
| CSI-RS | 32 ports, density 1 RE/port/PRB, 272 PRBs, 1 occasion / 20 ms | 8,704 | 0.47% | 0.69% |
| TRS | 52 PRBs, 2-slot occasion, 6 RE/PRB/slot, 4 beams, 2 occasions / 20 ms | 4,992 | 0.27% | 0.40% |
| Total | DDDSU; S slot with 6 downlink symbols, 4 guard symbols, and 4 uplink symbols | **234,112** | **12.76%** | **18.61%** |

**Table 3: Illustrative overhead breakdown for a 100 MHz 5G NR TDD carrier with DDDSU.**

### *C. Synchronization and System Information*

Initial access remains a delicate part of MRSS design because, unlike ordinary scheduled data, synchronization and basic broadcast functions cannot be moved around by the scheduler freely. In a pure 6G carrier, these functions would be designed entirely for 6G performance. In an MRSS carrier, however, they must also remain compatible with incumbent 5G operation and must ideally remain invisible to legacy 5G UEs. The latter would prevent 5G UEs from detecting or reporting 6G cells that would otherwise interfere with ordinary cell selection and mobility behavior.

The 6G SSB transmission must either use unused 5G SSB occasions or be protected by ensuring that 5G transmissions are not scheduled on the same resources. The same principle applies to system information and paging. In 5G, system information block (SIB) transmissions are dynamically scheduled like ordinary data, which is one reason their overhead can remain small. As long as the new signals remain hidden from 5G devices and occupy a small, predictable resource footprint, the coexistence burden of initial access can be kept very low.

### *D. Control Coexistence and CSI Reporting*

Downlink control channel is a critical MRSS design foundation. If 5G and 6G downlink control are not multiplexed efficiently, the shared carrier can lose a large fraction of its resources even though 5G is leaner than LTE. For example, Table 3 provides an illustrative example of overhead breakdown for a 100 MHz 5G NR TDD carrier with 30 kHz SCS and a DDDSU slot format. Table 3 is calculated over a 20 ms observation window. The carrier is approximated by 273 PRBs, and 20 ms contains 40 slots. Therefore, the total RE budget is $273 \times 12 \times 14 \times 40 = 1{,}834{,}560$ REs. For the DDDSU pattern, there are 32 downlink-bearing slots in 20 ms. As an example, the regular PDCCH overhead is calculated as $270 \times 12 \times 2 \times 32 = 207{,}360$ REs, corresponding to $207{,}360/1{,}834{,}560 = 11.30\%$ of total REs. The total downlink-bearing RE budget is $24 \times 273 \times 12 \times 14 + 8 \times 273 \times 12 \times 6 = 1{,}257{,}984$ REs, so the same PDCCH overhead corresponds to $207{,}360/1{,}257{,}984 = 16.48\%$ of downlink REs. It can be seen that the recurring regular PDCCH region is the dominant overhead source (~88.5%), while SSB, CORESET 0, SIB1, CSI-RS, and tracking reference signal (TRS) contribute much smaller overhead (~11.5%). This is why PDCCH design is at the heart of MRSS technical discussion in 3GPP. In 5G, PDCCH resides within configurable CORESETs, and within each CORESET the candidate locations are determined by a pseudo-random hashing function. For MRSS, several coexistence modes are possible. The most desirable mode is fully overlapping control, where 5G and 6G share the same CORESET region and perhaps even similar hashing logic. This would minimize overhead impact on an MRSS carrier due to 5G downlink control region.

One likely 6G rollout model would use a shared FR1 carrier as a coverage anchor and then aggregates it with one or more higher-band carriers for capacity. In such an architecture, the shared low-band uplink may carry not only UCI for the shared carrier but also CSI and feedback associated with multiple higher-band downlink carriers. The burden of reserved PUCCH resources and periodic CSI reporting can become significant, especially in small-bandwidth anchors. The burden could become larger when CSI for carrier aggregation was also mapped there. Therefore, it is necessary to study how 5G and 6G uplink control can be multiplexed and how the overhead impact of 6G uplink control reporting can be minimized when carrier aggregation is active. One potential idea is to move CSI reporting to PUSCH only, rather than reserving a large fixed PUCCH footprint.

### E. Reference Signals and Interference Awareness

A final core foundation of MRSS is explicit awareness of incumbent 5G signals. In MRSS, the relevant incumbent structures are 5G SSB, CSI-RS, and control-related resources. It is necessary to study interference caused by 5G common channels and reference signals to 6G UEs and investigate how to increase the awareness of 6G UEs operating in an MRSS carrier about those 5G signal patterns. The awareness could support several different mechanisms. One is rate matching, where resources coinciding with known incumbent structures are marked unavailable to the new RAT. Another is coordinated scheduling, where the scheduler uses shared knowledge of incumbent signal periodicity and footprint to avoid harmful overlap dynamically. A third is UE receiver-side mitigation, such as interference cancellation. Nonetheless, the DSS lesson was that leaving this entirely to advanced receiver implementations creates uncertainty and uneven deployment outcomes.

In summary, these foundations define MRSS as a co-designed shared-carrier framework. This is why MRSS should be understood not as a single 3GPP feature, but as a coordinated set of technical choices that together determine whether 5G-6G spectrum sharing can be efficient enough to serve as the main migration path. However, it should be noted that MRSS is a migration mode, not the pure-6G end state. This means that coexistence choices made for shared operation should not unnecessarily constrain how 6G evolves on pure 6G carriers. For example, reuse of 5G structures should be considered where it materially reduces coexistence burden, but not if it permanently compromises standalone 6G performance. This principle is critical because many carriers will eventually operate as pure 6G deployments. The coexistence framework must therefore be strong enough to support long migration periods, but clean enough that 6G performance does not remain artificially tied to incumbent assumptions once sharing is no longer needed.

## VI. CONCLUSIONS

The transition from DSS to MRSS reflects a broader shift in how the industry approaches spectrum migration. DSS demonstrated that dynamic sharing can accelerate the introduction of a new RAT in valuable incumbent spectrum, but it also revealed the practical costs of coexistence when the legacy system carries heavy fixed overhead, rigid control structures, and structured interference. Those lessons are relevant to 5G-6G migration. MRSS starts from a stronger technical foundation because 5G is much leaner and more configurable than LTE, and because 6G can be designed from the outset with coexistence in mind. Even so, efficient sharing will not happen automatically. The real success of MRSS will depend on whether the first 6G release can minimize overhead, support shared or efficiently multiplexed control, and provide sufficient awareness of incumbent 5G signals.

From a 3GPP perspective, the central question is no longer whether 5G and 6G can share spectrum, but whether they can do so efficiently enough to make shared FR1 operation the preferred migration path. If that challenge is met, MRSS can become an attractive framework for the 5G-to-6G transition, enabling continued reuse of scarce spectrum assets while preserving a clean path toward pure 6G deployments.

## REFERENCES

[1] H. Flinck *et al.*, “An overview of flexible spectrum management approaches for 6G," *IEEE Access*, vol. 13, pp. 207396-207411, 2025.

[2] 3GPP TR 38.914, “Study on 6G scenarios and requirements,” Version 0.4.0, Mar. 2026.

[3] W. Chen *et al.*, “5G-Advanced towards 6G: Past, present, and future,” *IEEE Journal on Selected Areas in Communications,* vol. 41, no. 6, pp. 1592-1619, Jun. 2023.

[4] R. K. Saha and J. M. Cioffi, “Dynamic spectrum sharing for 5G NR and 4G LTE coexistence - A comprehensive review,” *IEEE Open Journal of the Communications Society*, vol. 5, pp. 795-835, 2024.

[5] X. Lin, “A tale of two mobile generations: 5G-Advanced and 6G in 3GPP Release 20,” *IEEE Communications Standards Magazine*, 2026.

[6] Nokia, “Simplifying spectrum migration from 5G to 6G,” White Paper, 2023. [Online]. Available at https://www.nokia.com/asset/213378/

[7] S. Parkvall and Z. Du, “Multi-RAT spectrum sharing (MRSS) for efficient 5G-6G coexistence,” Ericsson Blog, Apr. 2026. [Online]. Available at https://www.ericsson.com/en/blog/2026/4/5g-6g-spectrum-sharing-mrss

[8] R1-2603175, “Operator views on MRSS,” Bouygues Telecom, Vodafone, Orange, Telecom Italia, British Telecom, Deutsche Telekom, AT&T, 3GPP RAN1#124bis, Apr. 2026. [Online]. Available at https://www.3gpp.org/ftp/tsg_ran/WG1_RL1/TSGR1_124b/Docs/R1-2603175.zip

[9] Y.-P. E. Wang *et al.*, “A Primer on 3GPP Narrowband Internet of Things,” *IEEE Communications Magazine*, vol. 55, no. 3, pp. 117-123, March 2017.

[10] A. Hoglund *et al.*, “Overview of 3GPP Release 14 further enhanced MTC,” *IEEE Communications Standards Magazine*, vol. 2, no. 2, pp. 84-89, Jun. 2018.

[11] MediaTek, “Dynamic Spectrum Sharing (5G),” White Paper, 2020. [Online]. Available at https://mediatek-marketing.files.svdcdn.com/production/documents/Dynamic-Spectrum-Sharing-WhitePaper-PDFDSSWP-0220-V3.pdf?dm=1684470662

[12] 3GPP TS 38.214, “NR; Physical layer procedures for data,” Version 19.3.0, Mar. 2026.

[13] A. Ghosh *et al.*, “5G evolution: A view on 5G cellular technology beyond 3GPP release 15,” *IEEE Access*, vol. 7, pp. 127639-127651, 2019.

[14] X. Lin *et al.*, “5G new radio: Unveiling the essentials of the next generation wireless access technology,” *IEEE Communications Standards Magazine*, vol. 3, no. 3, pp. 30-37, Sep. 2019.

[15] K. Takeda *et al.*, “Understanding the heart of the 5G air interface: An overview of physical downlink control channel for 5G new radio,” *IEEE Communications Standards Magazine*, vol. 4, no. 3, pp. 22-29, Sep. 2020.